# Centre for Technology Management Working Paper Series

# Crisis-Critical Intellectual Property: Findings from the COVID-19 Pandemic

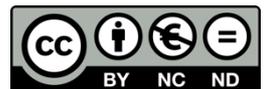

Frank Tietze [1,*], Pratheeba Vimalnath [1], Leonidas Aristodemou [1], Jenny Molloy [1,2]

[1] Innovation and Intellectual Property Management (IIPM) Laboratory, Centre for Technology Management (CTM), Institute for Manufacturing (IfM), Department of Engineering (CUED), University of Cambridge

[2] Open Bioeconomy Laboratory, Department of Chemical Engineering and Biotechnology (CEB), University of Cambridge

*Please contact the corresponding author for feedback:    frank.tietze@eng.cam.ac.uk

# Crisis-Critical Intellectual Property: Findings from the COVID-19 Pandemic

Frank Tietze[1,*], Pratheeba Vimalnath[1], Leonidas Aristodemou[1], Jenny Molloy[1,2]

*Abstract* — **Within national and international innovation systems a pandemic calls for large-scale action by many actors across sectors, in order to mobilise resources, developing and manufacturing Crisis-Critical Products as efficiently and in the huge quantities needed. Nowadays, this also includes digital innovations ranging from complex epidemiological models, artificial intelligence (AI) methodologies, to open data platforms for prevention, diagnostic and treatment.**

**Amongst the many challenges during a pandemic, innovation stakeholders and manufacturing firms particularly find themselves suddenly engaged in new relationships, possibly even with firms that have been competitors prior to the pandemic. Those stakeholders are thus likely to face intellectual property (IP) related challenges. Unfortunately, to (governmental) decision makers these challenges might not appear to be of paramount urgency compared to the many, huge operational challenges to deploy urgently needed resources. However, if IP challenges are considered too late, they may cause delays to urgently mobilising resources effectively. Manufacturing firms could be reluctant to fully engage in the development and mass manufacturing of Crisis-Critical Products.**

**This paper adopts an IP perspective on the currently unfolding COVID-19 pandemic to identify pandemic related IP considerations and IP challenges. The focus is predominantly on individual challenges and technical aspects related to research, development and urgent upscaling of capacity to manufacture Crisis-Critical Products in the huge volumes suddenly in demand. Its purpose is to provide a structure for those concerned with steering clear of IP challenges to avoid delays in fighting a pandemic.**

**From an ad-hoc patent analysis we identify that the majority of coronavirus related patents in the field are around organic chemistry, and development of methodologies and drugs for prevention, diagnosis and treatment of viruses. We also identify a time-lag between the outbreak and the materialisation of patent applications, which is consistent with the processes of the Patent Office. The large number of references to non-patent literature published after outbreaks is also an indication of the urgency of scientists to put the information in the public domain and make them accessible quickly to a wider audience.**

**We identify four stakeholder groups that are particularly concerned with IP related challenges during a pandemic. These include (i) governments, (ii) organisations owning existing Crisis-Critical IP (incumbents in Crisis-Critical Sectors), (iii) manufacturing firms from other sectors normally not producing Crisis-Critical Products suddenly rushing into Crisis-Critical Sectors to support the manufacturing of Crisis-Critical Products (new entrants) in the quantities that far exceed incumbents' production capacities and (iv) voluntary grassroot initiatives that are formed during a pandemic, often by highly skilled engineers and scientists to contribute to the development and dissemination of Crisis-Critical Products.**

**This paper discusses IP challenges faced by those stakeholders during a pandemic related to the development and manufacturing of technologies and products for (i) prevention (of spread), (ii) diagnosis of infected patients and (iii) the development of treatments. We offer an initial discussion of potential response measures to reduce IP associated risks among industrial stakeholders during a pandemic.**

## I. BACKGROUND

### A. The COVID-19 Pandemic

In December 2019, an outbreak of a novel coronavirus in Wuhan, Hubei province, manifested itself as a global health tragedy. The World Health Organization (WHO) announced it as a public health emergency of international concern on January 30, 2020 [1] and as a pandemic on March 11, 2020 [2]. The virus, later named SARS-CoV-2 [3], can cause mild flu-like symptoms (or even be asymptomatic) but can progress to acute pneumonia-like respiratory illness called novel coronavirus–infected pneumonia (NCIP). The overall clinical syndrome is known as COVID-19 [4]. Until today, there are no vaccines or medical cure for the disease yet [5] and the disease

---

*Corresponding author
*Emails*: Frank Tietze (frank.tietze@cam.ac.uk), Pratheeba Vimalnath (pv302@cam.ac.uk), Leonidas Aristodemou (la324@cam.ac.uk), Jenny Mollroy (jcm80@cam.ac.uk)

*Affiliations:*
[1] Innovation and Intellectual Property Management (IIPM) Laboratory, Centre for Technology Management (CTM), Institute for Manufacturing (IfM), Department of Engineering (CUED), University of Cambridge
[2] Open Bioeconomy Laboratory, Department of Chemical Engineering and Biotechnology (CEB), University of Cambridge

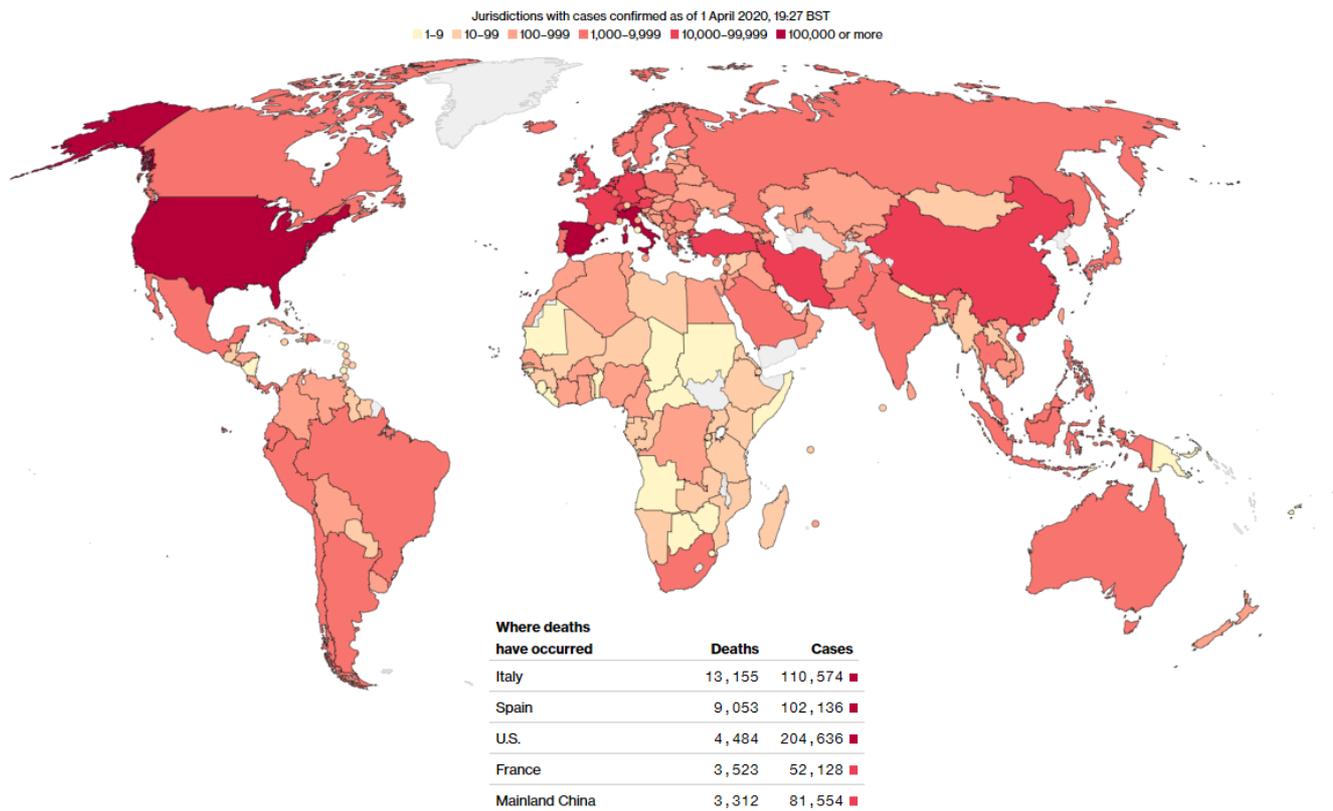

Fig. 1 COVID-19 spread (Source: Bloomberg, https://www.bloomberg.com/graphics/2020-coronavirus-cases-world-map/, 01.04.2020

has a fatality rate which is unconfirmed but likely to be around or above 1%.

According to the Centre for Systems Science and Engineering (CSSE), Johns Hopkins University & Medicine, as of April 1, 2020 there were about 911,308 confirmed cases worldwide, 45,497 deaths and 190,710 recovered cases (Fig. 1). Currently, over 180 countries are affected. Most countries have seen exponential growth rates in the number of cases with Italy (13,155), Spain (9,053), the United States (4,484) France (3,523), and Mainland China (3,312) being the most affected countries in terms of number of deaths, as of April 1, 2020 [6], [7]. The virus has a stronger transmission capacity than the 'conventional' annually recurring flu. On average, one infected person passes the virus to 2-2.5 others (that range is subject to change and can vary largely by geography, age group, and time) [8], [9].

The urgency of responding to the global threat of the pandemic has unleashed a plethora of efforts worldwide to tackle this pandemic as quickly as possible. Health care remains the utmost priority. Among other activities, research, technology and innovation efforts are pouring in to support health care systems through the development and ramping up the manufacturing of Crisis-Critical Products (CC-Products), such as Personal Protection Equipment (PPE), diagnostics tests, treatments, ventilators, vaccines etc.

While this calls for existing manufacturers (incumbents) that operate in Crisis-Critical Sectors (CC-Sectors) to quickly ramp up their production capacities, this also draws various other organisations into CC-Sectors that have not produced CC-Products prior to the pandemic. These 'new entrants' include large manufacturing firms from related and non-Crisis-Critical Sectors (CC-Sector), but also voluntary initiatives, not-for-profit organisations, scientists, engineers, universities, research institutions, start-ups and other forms of grassroots initiatives, etc., within the CC-Sector. Open Innovation efforts involving incumbents and new entrants become essential to address crisis-critical challenges. Incumbents typically own Crisis-Critical background IP (CC-IP) relevant for the manufacturing of CC-Products that new entrants lack. On the other hand when engaging in open innovation for the development and manufacturing of CC-Products, new entrants are likely to develop potentially valuable foreground CC-IP during the pandemic.

The purpose of this first paper is two-fold. First, we hope it contributes to raising awareness that IP considerations need to be addressed early rather than later during a pandemic. Second, we provide a structure (if not conceptual framework) for those concerned with steering clear of IP challenges, e.g. policy makers, governments, international organisations, large IP owners, new entrants, but also the many voluntary initiatives that are part of the grassroots movement.

This paper contributes to the many efforts to contain the pandemic as quickly as possible. We explore relevant IP considerations, provide a relevant terminology, describe the patent landscape of coronaviruses, systematically conceptualise three scenarios with different IP considerations and provide an initial discussion of possible approaches to reduce IP associated risks for relevant stakeholders, thus ensure that IP considerations do not delay the fight against a pandemic.

We apply an IP and innovation perspective on the enfolding pandemic and provide a systematic compilation, description and analysis of IP issues of relevance related to three critical areas for fighting the COVID-19 pandemic: (i) prevention of further spreading, including vaccine development, (ii) the development of diagnostics (i.e. test kits) for determining whether persons are infected or not and (iii) the (development of) treatments, i.e. a preventative-diagnostic-treatment framework.

B. Technology Related Pandemic Challenges

Applying an innovation and IP perspective we particularly focus on five technology related challenges that emerged from observations during the recent weeks of the pandemic, some relate to novel technologies highlighted in the WHO's Coordinated Global Research Roadmap for COVID-19 [10] while other have emerged from operations needs in frontline healthcare.

First and foremost, the challenge of finding a treatment for the acute respiratory pneumonia caused by COVID-19 has initiated large-scale R&D efforts. Second, the pandemic has created a sudden and massive demand for the development and manufacturing in extremely large volumes of diagnostic testing kits, not only with high accuracy that can be conducted in high capacity (e.g. for several weeks Germany alone has carried out 160,000 tests per week [11]) but also new ways of organising testing to be done (e.g. COVID-19 isolation pods, drive through testing). Third, the pandemic caused a sudden need to treat a large number of patients in hospitals requiring an extraordinarily large ICU capacity, particularly with an enormous need for certain medical devices particularly ventilator capacity (e.g. UK ventilator challenge [12]) by far exceeding the currently available capacity in many hospitals and countries. Fourth, the pandemic has called for a need of digital innovation, including epidemic modelling to monitor and understand the spread and development of the virus across populations, including tracking of cases and spreaders. Fifth, the COVID-19 pandemic has caused an exceptional high demand for skilled medical staff, doctors and nurses, particularly with ICU experience, such as anaesthetists and critical care nurses, all of whom need to be equipped with PPE (personal protective equipment), in this pandemic particularly protective clothing, face shields, goggles, gloves to protect health care staff from infection.

Additional challenges not addressed in this paper include the security of supply chains for essential goods; the impact of drastically reduced passenger and cargo transport routes and innovating at a health systems and infrastructure level to cope with testing and treating huge numbers of people. Also, the securing of food supply (chains) with supermarket chains play a major role, including the optimisation of delivery route planning, quick adjustments to online booking systems, e.g. rationing certain goods or prioritising delivery slots to the elderly and vulnerable. Other challenges then relate to the Information Technology (IT) and infrastructure for connecting all the people suddenly trying to work at home, i.e. video conferencing platforms and equipment (e.g. Google and Microsoft announced free access to their advanced tele-conferencing and collaboration tools [13] as well as internet service providers relieving data caps [14]). And then there are urgent logistical challenges, for instance, to efficiently reorganise supply of Crisis-Critical Products (CC-Products), the repatriation of national citizens stuck abroad, but also internal operations processes in hospitals as wards have been repurposed and specific COVID-19 testing pods have been setup. This list is obviously not exhaustive.

C. Innovative and Technology Responses to the Pandemic

During the past weeks we have observed a number of responses to the five challenges outlined above. Pharmaceutical, biotechnology firms and universities have joined forces to develop vaccines [15] and treatments [16], also testing whether existing antiviral drugs could be repurposed, e.g. malaria / HIV drugs or the development of novel COVID-19 specific drugs [17]. For instance, a collaboration of Clover and GSK has been announced recently [18]. Another consortium includes life big Pharmaceutical companies such as Novartis, Bristol Myers Squibb and GSK [19]. Others started to develop novel diagnostic test, such as BOSCH recently having announced they had developed their own COVID-19 test kit [20]. Manufacturing companies from all kind of sectors have started to repurpose their production lines to support the production of Crisis-Critical Products, involving large engineering / manufacturing firms such as those involved in the UK ventilator challenge consortium (e.g. Airbus, GKN, Roll-Royce, Siemens, Smiths group) [12], [21], [22], but also luxury brands (e.g. French LVHM) using perfume manufacturing facilities to suddenly make hand sanitisers [23], as well as SMEs starting to produce sanitisers [24], textile manufacturers (e.g. ZARA in Spain [25], Trigema in Germany [26], Prada in Italy [27]) to produce face masks. Various volunteering initiatives started to emerge run by scientists and engineers to (i) develop open hardware / source designs of ventilators, (ii) find new ways of design and manufacture PPE, e.g. 3D printed face shields [28] and ventilator valves [29].

Last but certainly not least, digital innovations have sprung up widely, e.g. data / software approaches by scientists for prevention, diagnostic and treatment. Focusing on preventative digital innovation, scientists have focused more on open data platforms, by developing epidemiological models to understand government responses and forecast the growth curves of the virus [31], analysing geospatial models to understand the distribution and spread of the virus [32], and deploying causal-effect models to understand symptoms of the virus and limit its spread through behavioural science, and tracking applications [33]. In the diagnostic sphere, scientists utilise AI, and more specifically Deep Convolutional Neural Networks to detect COVID-19 from X-Ray Images. This has also been particularly useful in diagnostic analysis of symptoms to predict the development of a patient's case [34]. A high number of efforts have also been concentrated on treatments, where scientists have developed AI text and data mining tools to help the medical community develop answers to high priority scientific questions and potential treatments. Efforts have focused on the development and summarisation of genome specific medical protocols of precision medicine and host response, as well as modelling and simulation of the virus propagation and efficiency of interventions [35].

From what emerged during this pandemic in the news and expert discussions, one can categorize the crisis-critical activities in three categories, most of them related to innovation or massive capacity building / upscaling to manufacture and supply CC-Products in sufficient quantities in a short period of time. The first category is prevention, including digital innovations to track the virus spread, sanitisers, PPE equipment etc in order to slow down the spread of the virus, but also vaccines to control future outbreaks which are currently under development. The second category is diagnostics predominantly including the need for an incredible volume of testing kits, and those that are accurate, but also portable and deliver speedy results, recently also antibody tests. The third category is treatment, including development of treatments through repurposing or existing drugs, development of new antiviral drugs, but also ventilators for ICU critical care in hospitals around the world. Those three categories are used throughout the remainder of this paper.

## II. Why worry about IP during a Pandemic

A general characteristic of IP is that it is intangible and remains largely invisible in most day to day operations of companies and economies. Yet, it is vital for the functioning of today's economy in normal but likewise in times of crisis.

For the purpose of this paper we define "Crisis-Critical IP" (CC-IP) as the set of intellectual property, relevant for the research, development, manufacturing and distribution of Crisis-Critical Products, Services and Technologies that are urgently needed for quickly ending a crisis situation, with crisis defined as a situation, that threatens the human species and an approved world body elevates the risk as critical, i.e. in the case of COVID-19 the WHO has declared this situation a pandemic. CC-IP includes (i) formal and registered IP, such as patents, design rights, trademarks, (ii) formal unregistered IP, e.g. copyrights, design drawings, CAD files, trade secrets for manufacturing processes, as well as (iii) informal IP, such as know-how. CC-IP refers both to already existing (background) IP predominantly owned by incumbents that already operate in CC-Sectors prior to a crisis as well as novel (foreground) CC-IP that is developed during the crisis by various kind of actors, including incumbents, but also new entrants to CC-Sectors.

Certain CC-Sectors and its related technologies, products and services have different degrees of formal and informal IP. Some of these products are more 'high-tech' than others. Previous studies show that IP appears to be of particular importance in certain sectors, such as electronics, ICT, high value manufacturing, software, but also pharmaceutical, biotechnology, medical devices and life sciences. Many of those IP intensive sectors are of particular relevance in today's unfolding pandemic. For instance, ventilators are typically expensive medical devices and the incumbent manufacturing companies are likely to own alive / active (i.e. not yet expired) CC-IP, PPE equipment (e.g. face masks) being fairly 'low tech', with a high probability that formally relevant CC-IP has expired.

While IP issues have hardly surfaced during the beginning of this pandemic, this has happened quite recently. A few examples are summarised below. The Wellcome Trust appears to be among the first prominent organisations that understood the relevance of IP for this pandemic early on [36]. With a particular focus on research, on January 31, 2020, the trust called for journals, publishers etc. to allow widespread sharing of all potentially relevant research and dataset. This initiative is geared towards publishers to not put any COVID-19 relevant publications behind a paywall. The pledge seems to be a huge success as a wide range of renowned organisations have now signed up to it, including leading journals, such as Nature and The Lancet, but also the European Commission, publishers (e.g. Cambridge University Press), national academies of science (e.g. Academy of Medical Sciences, The Royal Society), foundations (e.g. Bill & Melinda Gates Foundation), research councils (e.g. UK Medical Research Council), ministries (e.g. Indian Department of Biotechnology, Ministry of Science & Technology) and a wide range of other organisations, including companies (e.g. BenevolentAI, Johnson & Johnson). By now more than 24,000 research papers are available online [37]. In the past weeks some other organisations have started raising concerns that IP might be an issue during the pandemic and have called for the government and private sector to respond. For instance, on March 27, 2020, Doctors Without Borders (MSF) publicly announced their concern that firms might try to profiteer from the crisis [38] and the government of Costa Rica called the WHO to organise the pooling of relevant IP [39]. By now, a few governments have passed compulsory license resolutions for CC-IP, e.g. Chile and Canada [40] and some even authorized issuance of compulsory license e.g. Israel's compulsory licensing for Kaletra [41]. Compulsory licensing schemes are typically some kind of "last resort" government measures.

We have also observed some initial approaches that firms have taken related to IP during this pandemic. For instance, measures to mitigate the risk of counterfeit products being distributed in the crisis, e.g. PPE masks [42], [43]. Some companies enforced IP lawsuits against other companies developing CC-Products. For instance, Labrador Diagnostics LLC sued BioFire, a company developing COVID-19 testing kits for infringing two of its patents [44], but later announced a royalty free licensing to anyone developing tests [45]. Some first firms have already filed patents or other forms of exclusivity, e.g. Gilead applied for "orphan drug" designation, but dropped it a few days later after criticism [46], Pharma firms filed for repurposing of drugs extending indications e.g. Remdevir [46], [47].

Very recently we have seen a limited number of firms adopting at least some selective open IP approaches, particularly pledging [48] relevant IP, such as Fortress [49], AbbVie and medical device companies manufacturing ventilators sharing IP (including design specifications and files), such as Irish Medtronic [50] and UK based Smiths Group [51]. A recent initiative of scientists and lawyers has launched the Open COVID Pledge (www.opencovidpledge.org) calling IP owners to not assert relevant IP during the crisis until one year after the WHO declares the pandemic to be over [52]–[54].

### III. CORONAVIRUS PATENT ANALYSIS

One of the main IP challenges, both in this pandemic, and in general is the availability of open data for analysing the progression of the virus [55], as well as the different analysis types deployed [56]. In an outbreak so severe as the COVID-19, where the reported cases to this moment are close to 1 million worldwide, any available dataset is potentially helpful to derive insights into the disease. We perform an ad-hoc patent analysis for coronaviruses using an open dataset by Lens.org[1], to enhance our understanding into preventive, diagnostic and treatment measures for the virus. We focus on the broader spectrum of coronaviruses to identify patterns from earlier outbreaks that could be applied in the case of COVID-19. Table 1 shows the descriptive statistics and correlations for the patent dataset and Fig.2 shows some initial results from the Coronavirus Patent Analysis.

Fig. 2a shows the top 10 CPC classification distribution at the subgroup level. It is evident that the highest number of patents belonging to the primary CPC subgroup classification are in C07K14/005. This is the organic chemistry subclass, for peptides with more than 20 amino acids and specifically for viruses, which constitute viral proteins. This is followed by the subgroups A61K39/215 (medicinal preparations containing antigens or antibodies materials for immunoassay for coronavirus) and C12Q1/701 (measuring or testing processes involving enzymes, nucleic acids or microorganisms involving virus specific hybridization probes). These are followed by C12N7/00 (preparation of viruses bacteriophages compositions, medicinal viral antigen or antibody compositions), G01N1 (investigation processes for measuring or testing other than immunoassay, involving enzymes), C07K16/08 (investigation of immunoglobulins from RNA viruses). Within the top 10 classes, we also find a subclass within section Y (Emerging Cross-Sectional Technology), with Y02A50/451, which is specific for genetic or molecular screening of pathogens. This indicates that within the granted patents, there are some vaccination related patents available. All the top 10 CPC classifications have to do specifically with the chemical characteristics of the virus for prevention, diagnostics and treatment. Also, one could note that granularity of patent applications in these fields since some of these have 50 CPC subgroups.

Fig. 2b shows the top 10 IPC classification distribution at the subgroup level. There are 103 unique subgroups referenced within these patents. While the highest primary IPC sub group is A61K39/215 (preparation for medical purposes devices or methods specially adapted for bringing pharmaceutical products into particular physical or administering forms chemical aspects of, or use of materials for deodorisation of air, for disinfection or sterilisation, or for bandages, dressings, absorbent pads or surgical articles soap compositions for coronavirus), the highest collective number is for the subgroup, C12N15/09 (microorganisms or enzymes compositions thereof propagating, preserving or maintaining microorganisms mutation or genetic engineering culture media microbiological testing media recombinant DNA-technology).

Moreover, 64% of the top 10 IPC subgroups, fall within the A61K subclass [57], which covers the following subject matter under a mixture composition or process of preparing a composition or treating process: drug or other biological compositions which are capable of: preventing, alleviating, treating or curing abnormal or pathological conditions of the living body by such means as destroying a parasitic organism, or limiting the effect of the disease or maintaining, increasing, decreasing, limiting, or destroying a physiological body function; diagnosing a physiological condition or state by an in vivo test; in vitro testing of biological material. The rest of the subclasses fall within the C12 class, where viruses, undifferentiated human, animal or plant cells, protozoa, tissues and unicellular algae are considered as microorganisms. From both Fig.2a and Fig.2b, it appears evident that the highest clusters of patents are around the chemical process of identification, composition and vaccine development for the coronavirus family.

Fig. 2c shows the distribution of patent applications within the different jurisdictions. Fig.2d shows the distribution of the granted patents within different jurisdictions. Immediately, evident is that the granted patents are lagging behind the patent applications. There are some old granted patents that fall within the broad spectrum of the analysis in 1975 about the development of vaccines, however patent applications before 1975 are not included in the dataset. The majority of patent applications are filled in the US, followed by the transition to WO under the Patent Cooperation Treaty (PCT). These are followed by AU, CN, EP, CA, KO and JP. The distribution of granted patents follows an upwards trend (greater rate of increase than that of patent applications), but with significantly lower numbers (with the peak around 120 granted patents in 2019 relative to an average of 300 patent applications from Fig.2c). The distribution of these granted patents within the different jurisdictions are US, EP, AU, CN, KO, RU, and JP. It is also interesting to note that the conversion rate of patent applications to granted patents is 32%.

---

[1] Human Coronavirus Innovation Landscape: Patent and Research Works Open Datasets. Accessed 01.04.2020 at https://about.lens.org/covid-19

Table 1 Descriptive Statistics and Correlations for Coronavirus: Broad Keywords Based Patents (Patents = 6896, Patent Families = 2670)[2]

| | Publication Year | Application Year | Earliest Priority Year | Number of Applicants | Number of Inventors | Forward Citations | Simple Family Size | Extended Family Size | Sequence Count | No. of CPC Subgroups | No. of IPC subgroups | NPL Citation Count |
|---|---|---|---|---|---|---|---|---|---|---|---|---|
| **Mean** | 2010 | 2008 | 2007 | 1.99 | 3.78 | 4.78 | 13.81 | 20.58 | 465.78 | 5.97 | 5.88 | 5.62 |
| **Std. Error** | 0.08 | 0.08 | 0.08 | 0.02 | 0.03 | 0.20 | 0.17 | 0.57 | 156.48 | 0.06 | 0.08 | 0.23 |
| **Median** | 2011 | 2008 | 2006 | 1.00 | 3.00 | 0.00 | 9.00 | 11.00 | 0.00 | 3.00 | 4.00 | 0.00 |
| **Mode** | 2005 | 2004 | 2003 | 1.00 | 2.00 | 0.00 | 2.00 | 2.00 | 0.00 | 3.00 | 3.00 | 0.00 |
| **Std. Dev.** | 6.33 | 6.55 | 6.53 | 1.99 | 2.48 | 16.81 | 14.18 | 47.58 | 12994 | 5.20 | 6.20 | 19.27 |
| **Kurtosis** | 1.09 | 0.93 | 0.93 | 13.67 | 7.01 | 313.38 | 4.12 | 318.23 | 1354.76 | 9.65 | 19.00 | 66.04 |
| **Skewness** | -0.78 | -0.65 | -0.60 | 3.10 | 2.30 | 13.22 | 1.87 | 14.88 | 35.86 | 2.60 | 3.22 | 6.76 |
| **Min.** | 1975 | 1973 | 1972 | 0.00 | 2.00 | 0.00 | 1.00 | 1.00 | 0.00 | 3.00 | 0.00 | 0.00 |
| **Max.** | 2020 | 2020 | 2020 | 18.00 | 22.00 | 555.00 | 117.00 | 1173.00 | 544026.00 | 56.00 | 103.00 | 342.00 |
| **Correlations** | | | | | | | | | | | | |
| Publication Year | 1 | | | | | | | | | | | |
| Application Year | 0.94 | 1 | | | | | | | | | | |
| Earliest Priority Year | 0.91 | 0.96 | 1 | | | | | | | | | |
| Number of Applicants | -0.09 | -0.04 | -0.04 | 1 | | | | | | | | |
| Number of Inventors | 0.12 | 0.12 | 0.13 | 0.36 | 1 | | | | | | | |
| Forward Citations | -0.21 | -0.18 | -0.17 | 0.12 | 0.00 | 1 | | | | | | |
| Simple Family Size | -0.01 | -0.08 | -0.15 | -0.05 | -0.01 | 0.03 | 1 | | | | | |
| Extended Family Size | -0.05 | -0.07 | -0.11 | 0.03 | 0.03 | 0.12 | 0.34 | 1 | | | | |
| Sequence Count | 0.02 | 0.02 | 0.03 | 0.03 | 0.01 | 0.00 | -0.02 | -0.01 | 1 | | | |
| No. of CPC Subgroups | -0.10 | -0.10 | -0.13 | 0.12 | 0.05 | 0.14 | 0.00 | 0.05 | 0.04 | 1 | | |
| No. of IPC subgroups | -0.21 | -0.26 | -0.29 | -0.02 | 0.00 | 0.07 | 0.26 | 0.12 | -0.01 | 0.20 | 1 | |
| NPL Citation Count | 0.10 | 0.08 | 0.05 | 0.07 | 0.04 | 0.10 | 0.04 | 0.05 | 0.00 | 0.05 | 0.00 | 1 |

*Note: Search Query: (title:(Coronavirus) OR abstract:(Coronavirus)) OR claims:("Severe acute Respiratory syndrome") OR abstract:("Severe acute Respiratory syndrome") OR claims:("COVID 19") OR claims:("Severe acute Respiratory syndrome") OR (title:("coronaviridae") OR abstract:("coronaviridae")) OR claims:("SARS-CoV") OR claims:("MERS-CoV") OR claims:("COVID 19") OR claims:("Wuhan coronavirus") OR claims:("2019-nCoV") OR claims:("Middle East respiratory")*

[2] Human Coronavirus Innovation Landscape: Patent and Research Works Open Datasets. Accessed 01.04.2020 at https://about.lens.org/COVID-19

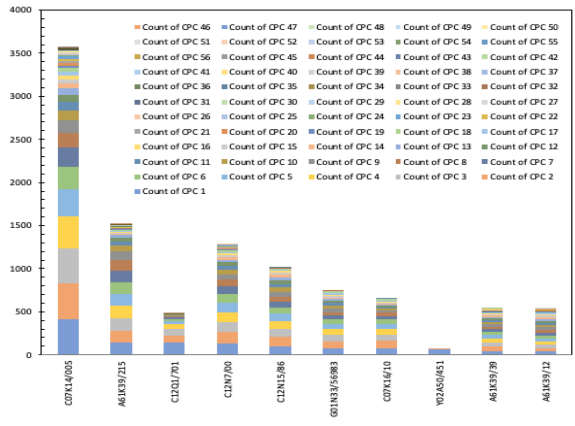

(a) Top 10 CPC Classification Distribution
(sorted by descending order of primary CPC main group)

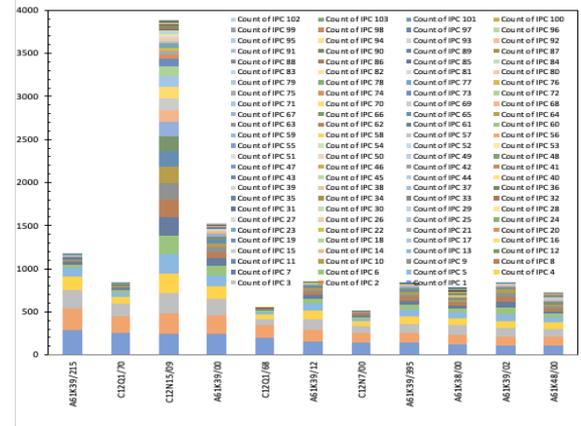

(b) Top 10 IPC Classification Distribution
(sorted by descending order of primary IPC main group)

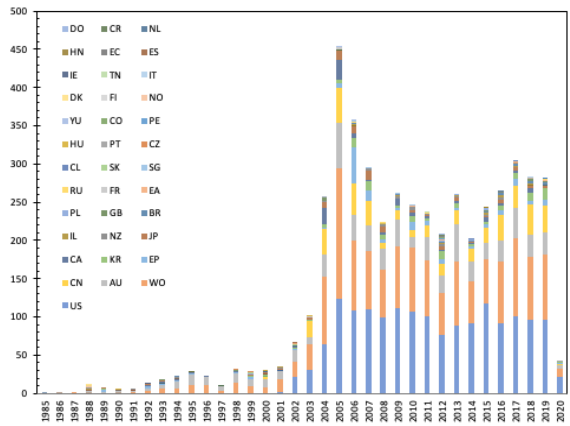

(c) Patent Applications Distribution vs. Publication Year
(filter by jurisdiction)

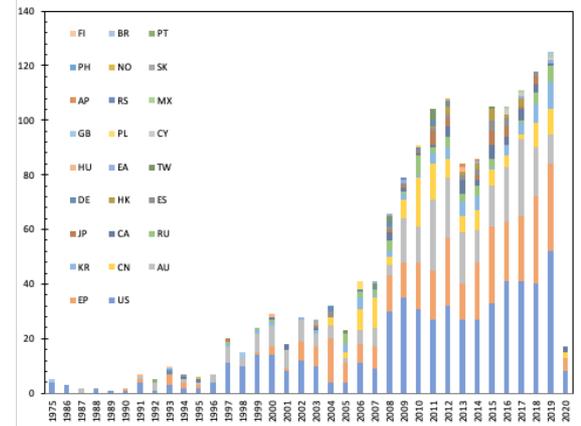

(d) Granted Patents Distribution vs. Publication Year
filter by jurisdiction

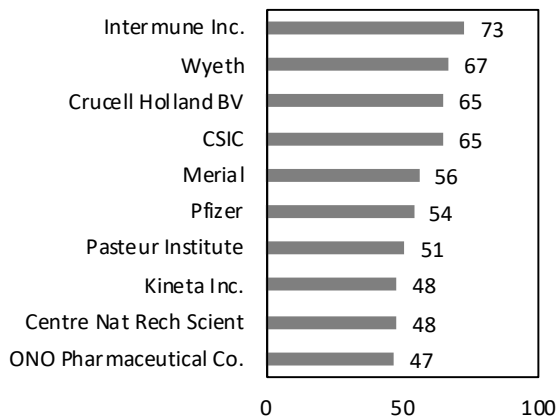

(e) Top 10 Applicants
(co-applications are also included in the above data)

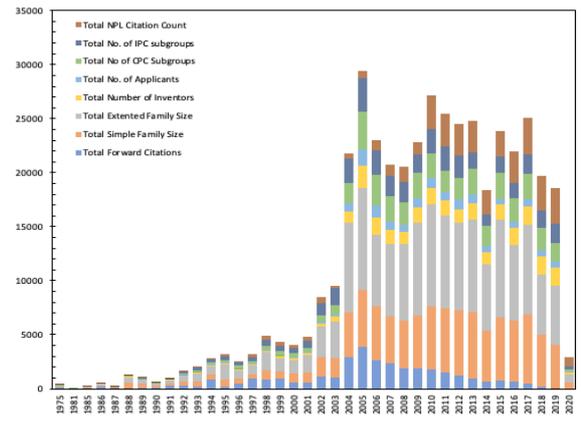

(f) Distribution of Patent Characteristics vs. the Publication Year

Fig. 2 Results from the Coronavirus Patent Analysis

## IV. IP Considerations during a Pandemic

What in normal circumstances are considered typical IP activities are no more normal during times of pandemics. In this 'new normal', our analysis about the IP related considerations that arise during the efforts to end the COVID-19 pandemic identifies scenarios with IP relevance involving four main stakeholder groups shown in

Our analysis finds that during the COVID-19 pandemic a range of non-Crisis-Critical Product (CC-Product) manufacturers entered Crisis-Critical Sectors (CC-Sectors), in which incumbent manufacturers have developed, produced and supplied Crisis-Critical Products (CC-Products) already before the pandemic. Those existing firms had insufficient production capacities to supply Crisis-Critical Products (CC-Products) in the huge quantities needed in a timely manner, leading to supply shortages for customers. Firms from non-Crisis-Critical Sectors (CC-Sectors), such as 3D printing, automotive, aerospace, home appliances, fashion and luxury goods, rushed into CC-Sectors to help cope with the CC-Product supply shortages

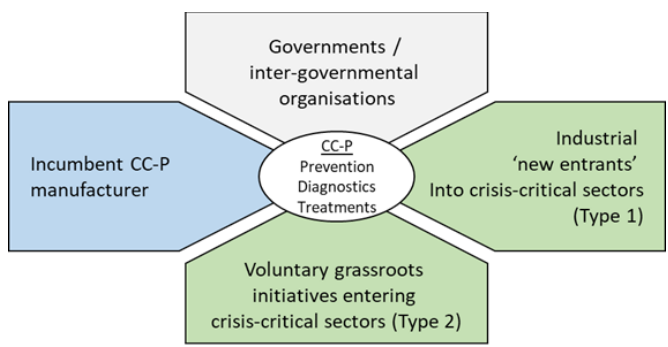

Fig. 3 Four Main Stakeholder Groups that are concerned with IP during a Pandemic

Table 2 provides an overview of the IP related considerations for COVID-19 (with examples) distinguishing (i) the prevention of COVID-19 (including measures to limit its spread and vaccines to prevent future outbreak), the (ii) diagnostics (including professional and self-testing) and (iii) treatment, with the latter including the direct treatments (e.g. development of drugs) but also the treatment of symptoms, i.e. related to the medical equipment needed to keep bodies alive (e.g. ventilators, ICU beds).

The following conceptual scenarios are built on the distinction between these groups of stakeholders. First, we label firms as 'incumbents' that already operated in CC-Sectors before the start of the pandemic developing, manufacturing and supplying CC-Products. These incumbents are highly likely to own CC-IP when a pandemic starts. Second, we label those organisations 'new entrants' that suddenly rush to enter the CC-Sectors after the beginning of a pandemic in order to support scaling up the development and manufacturing of CC-Products. New entrants can be industrial manufacturing firms (Type 1), but also voluntary grassroot initiatives, including not-for-profit communities, start-ups, entrepreneurial scientists, etc (Type 2). While manufacturing firms entering CC-Sectors are highly likely to own IP, this might not be relevant CC-IP. In contrast, voluntary grassroot initiatives typically do not own formal IP prior to a pandemic as these only form during a pandemic. Based on that distinction we identify three broad scenarios that are discussed below.

### A. Scenario 1: Type 1 New Entrants – Non CC-Product Manufacturers

Existing, often large manufacturing firms that did not produce CC-Products before the start of a pandemic are (i) either called in (ordered) to help with upscaling the production of CC-Products by governments (e.g. UK government's call for firms to produce ventilators [12], [92]) or (ii) voluntarily switch their production to CC-Products, e.g. because their usual products are not in demand during the pandemic (e.g. luxury company LVHM starting to produce health care products like sanitisers, hydroalcoholic gel [93]). These firms then become new entrants into CC-Sectors. There appears to be four kinds of new entrants that suddenly rush to help with producing CC-Products by repurposing their production lines.

First, despite coming from different sectors, certain specialized companies at least have manufacturing capacity as well as a set of valuable resources and capabilities that are somewhat related and can be readily used with a minimal change for production of certain CC-Products to meet the supply shortage. Examples include companies such as LVHM, which had been producing perfumes before the COVID-19 pandemic, so own manufacturing process equipment to fill bottles with alcohol containing liquid. During the pandemic they changed the liquid from perfume to sanitiser.

The second type of manufacturing companies are those that possess relevant expertise, resources and competence that can be put in use for developing and manufacturing CC-Product to address a shortage of certain CC-Products. An example is the UK company Dyson, which has long been in the air-flow business for its vacuum cleaners and hair dryers, has expertise in air purification technologies with some of its technologies like its digital motor already optimized for safety and efficiency. Dyson hence was able to develop a new ventilator design called CoVent in just 10 days [78]. Another example is GENTL masks, an open source design for masks by EPAM, an engineering and software solution provider and platform developer [65]

The third type includes technology giants that possess diverse capabilities and rich resources that they can deploy to basically manufacture any product using IP from incumbents in the CC-Sectors. Those include automotive companies entering the production of ventilators (e.g. Volkswagen [94]), but also multinationals like BOSCH starting to produce diagnostic test kits [20].

Table 2 IP Considerations for COVID-19, synthesized with an adapted Intellectual Property (IP) Roadmapping Framework

Table 2a: IP Considerations for Technologies, Products, Services, IP Ecosystem, Purpose and Challenges

| COVID-19 | Why? | Who and what? | | |
|---|---|---|---|---|
| | Technologies, products, services | IP ecosystem stakeholders | IP purpose | IP Challenges |
| | Which technologies are relevant? What types are these? How are important platforms developing? | Consider the different IP ecosystem actors. Which actors are likely to play important roles? Which roles do they play? | What purpose does IP Serve? | |
| **Prevention** | *Vaccine:* 1) Vaccines to prevent future outbreak (Table 3) [58]-[60]; 2) Tools to monitor the spread of the disease [61]; 3) Epidemiological models to forecast the growth curves) [31]; 4) Analysis of geospatial data of COVID-19 cases and 5) Causal Effects and in-depth analysis of COVID-19 causes, and effects, symptoms; Applying machine learning/AI methods to mitigate the spread of the COVID-19 [62] *Medical equipment / PPEs:* 1) Diagnostics kits ([63], [64] for list of approved rapid and other COVID-19 test-kits) 2) Masks to prevent disease spread ([67]-[69] using list of DIY masks) 3) Sanitizers to avoid spreading of disease 4) Specialized products to avoid spreading of disease [70] | *Incumbents from CC-Sectors: Medical technology & tool developers:* 1) Traditional health incumbents develop test-kits (Everywell [71]) 2) Digital health community develop digital tools to monitor spread [61] 3) Big pharma and drug development firms: GSK, J&J, Gilead, Zydus Cadila, [18], [75] *Medical and healthcare service providers:* 1) Testing service providers (telemedicine doctors), PWNHealth [71] *New entrants from non CC-Sectors:* 1) Universities (vaccines, sanitizers) (e.g., Oxford's vaccine [76], IIT Delhi's Sanitizer) *New entrants from CC-Sectors. (New entrant type 1):* 1) Engineering and software solution providers offer open source designs for masks (e.g. GENTL mask, open source by EPAM [65]) 2) IT giants give free platform access to support remote working [13] 3) 3D printing development, hands-free door handle attachments [70] *Start-ups from CC-Sector (New entrant type 2):* 1) Start-ups developing testing solutions & partnering with medical centres (blood testing by Sight Diagnostics, Israel start-up with Sheba Medical) [79] *New entrants from non CC-Sectors (New entrant type 1):* 1) Publishers providing free access to research and technologies (e.g. Springer Nature, PubMed Central, Association of American Publishers, IEEE [37], [80], [81]) *Charity creators/integrators:* 1) Platforms to pledge IP (e.g., Open Covid pledge) [52]-[54] *Government bodies:* 1) WHO (e.g. provides standard for sanitizers), asked to create patent pools); 2) National governments (announcing and considering compulsory licensing [74]; 3) Government-industry engagements (e.g. Johnson & Johnson expanded an existing R&D agreement with the US Department of Health and Human Services (HHS) to develop COVID-19 vaccine) [83] | 1) Addressing supply shortages of Masks. Free design / DIYs free up the demand for clinical grade masks for use in hospitals [66] 2) Access to critical and relevant knowledge, and social awareness through copyrights waiver and offering free contents [37], [80], [81] 3) Accelerating vaccine development through consortiums and IP pooling and expertise [84] | IP imitation and risk of counterfeit products [84] |
| **Diagnostics** | *Medical equipment/ methods:* 1) Testing kits (Self-testing kit/ Professional testing kit): and chloroquine and hydroxychloroquine 2) Home testing-kit [71] 3) Rapid testing technology (e.g. [63], [64] for list of approved rapid and other COVID-19 test-kits) [73] 4) AI tools - Detection of COVID-19 using Deep CNNs from X-Rays [34] 5) Diagnostic analysis and forecasting of symptoms [73] | *Incumbents from CC-Sectors: Medical technology & tool developers:* 1) CC-Sector incumbents 2) Universities (E.g., Oxford's rapid testing technology) [72] *Medical and healthcare service providers:* 1) Testing service providers (telemedicine doctors), PWNHealth [71] | Accelerated development and wider availability of testing kits through IP sharing [35] | 1) Patent infringement lawsuits attempts delaying development of testing by existing players (e.g., lawsuit by Labrador Diagnostics LLC against BioFire) [44] |
| **Treatment** | *Drugs:* 1) Existing drugs that can potentially treat COVID-19 (repurposing) - remdesivir, lopinavir and ritonavir in combination; lopinavir/ ritonavir plus interferon-beta; 2) Genome-specific COVID-19 medical protocols, including precision medicine [74] 3) Modeling, simulation, of COVID-19 propagation and efficacy of interventions [62] 4) Design and sharing of clinical trials for analysis on medications, and therapies [35] 5) Develop AI text and data mining tools that can help the medical community develop answers to high priority scientific questions about treatments [35] *Medical equipment:* 1) Ventilators (lightweight / portable) valves [22],[29] 2) Medical grade masks [27] | *Incumbents from CC-Sectors: Drug developers:* 1) Big pharmaceutical companies: GSK, Novartis, Bristol Myers Squibb, AbbVie [19] *Medical technology & tool developers:* 1) Digital health community with tools for treatment [61] *Universities:* 1) University start-ups make portable ventilators (e.g. University of Toronto) [77] *New entrants from non CC-Sectors:* 1) 3D printing companies prototyping and manufacturing ventilators and valves [29] 2) Aerospace and automotive industry into ventilator manufacturing [12], [21], [22] 3) Household appliances industry (e.g. Dyson's CoVent applying Dyson's air purifier expertise to ventilators) [78] 4) Fashion industry for medical masks (e.g. Zara [25], Trigema [26], Prada [27]) | 1) Wider accessibility of medicines through access to patents (e.g., AbbVie has agreed to drop enforcement of Kaletra patents worldwide) [85] 2) Addressing supply shortages in ventilators, Ventilator Challenge UK consortium [22] 3) Accelerated development and wider availability, open initiatives for ventilators [86] | 1) Declining access to existing Crisis-Critical IP may lead to reverse - engineering by new entrants from non CC-Sectors (e.g., Italian volunteers 3D print reverse-engineered valves after denial of IP) [29] 2) Patents as barriers to the production and provision of low-priced treatments [74] |

Monopoly over CC-IP may increase production cost of the COVID-19 related medicines and in turn the price. Proactive measures by governments is the implementation of compulsory licensing [74]

Table 2b IP Considerations for IP Assets and IP Strategies

| COVID-19 | | Prevention | Diagnostics | Treatment |
|---|---|---|---|---|
| **IP Assets and Strategies** | **IP Assets** | Designs (DIY masks, hand free door handle attachment) [66],[70] | Patents for testing technology | Patents (e.g., drug related patents including product and process patents, patents of technologies for medical equipment like ventilators, valves); New patents (e.g., Isinnova plans to patent Charlotte Valve) [29] |
| | | Patents (e.g. vaccination related) | Designs for test-kits | |
| | | Copyrights - free copyright (open access articles) provides access to relevant, knowledge and research which would otherwise be not available [37], [80], [81] | | |
| **How?** | Which IP strategies are best suited to help achieve the IP purpose? | 1) Open source design for masks. (e.g., GENTL mask, an open source design by EPAM [65]) | 1) Royalty free licensing of patent-protected diagnostics technology (e.g., Labrador Diagnostics, a subsidiary of Fortress Investment Group) [45] | Compulsory licensing: 1) Compulsory licensing for COVID-19 treatment drugs (e.g., Canada, Chile, Israel) [74] |
| | | 2) Free access for limited time (e.g., IT giants Google & Microsoft providing free access to their conference tools to support remote working) [13] | | Free access: 2) Obtaining new patent protection and giving the patent free for others to use [88]; free access and usage rights to existing patents [85], [50] |
| | | 3) Bilateral collaborations for vaccine development (e.g., Sanofi with Translate Bio; GSK with Clover) [18],[87] | | Multi sector collaboration: 3) Multi-sector consortia to accelerate ventilator production to address demand-supply shortage. E.g., Ventilator Challenge UK consortium involving aerospace, automotive and medical industry players [22] |
| | | 4) Consortia for vaccine R&D [84] | | 4) R&D in non-CC-Sectors to apply expertise from non CC-Sectors to develop medical equipment (e.g., Dyson's CoVent applying Dyson's air purifier expertise to develop ventilators) [78] |
| | | 1) Voluntary pool by governing bodies (e.g., WHO asked to create voluntary pools [39]) | | |
| | | 2) Consortia among big pharma companies to accelerate the development, manufacture and delivery of vaccines, diagnostics, and treatments for COVID-19' [89] | | |
| | | 3) Pledge to give free access to all CC-IP for limited time. Voluntarily or through third-party platforms (e.g., Open Covid pledge) [52], [53] | | |
| | | 4) Hackathons to share and generate ideas [90] | | |
| | | 5) Open Platform Sharing of Datasets and Source Code [91] | | |

Table 3 Vaccine Development Sample, total of 54 Vaccines (2 in phase 1 clinical trials and 52 in pre-clinical)*

| | Platform | Type of Candidate Vaccine | Developer | Clinical Evaluation Stage | Platform (non-Coronavirus) |
|---|---|---|---|---|---|
| 1 | Non- Replicating Viral Vector | Adenovirus Type 5 Vector | CanSino Biological Inc. and Beijing Institute of Biotechnology | Phase 1 ; ChiCTR2000030906 | Ebola |
| 2 | RNA | LNP- encapsulated mRNA | Moderna/NIAID | Phase 1 ; NCT04283461 | multiple candidates |
| 3 | DNA | DNA plasmid vaccine Electoporation device | Inovio Pharmaceuticals | Pre-Clinical | Lassa, Nipah, HIV, Filovirus, HPV, Cancer indications, Zika, Hepatitis B |
| 4 | DNA | DNA | Takis/Applied DNA Sciences/Evvivax | Pre-Clinical | |
| 5 | DNA | DNA plasmid vaccine | Zydus Cadila | Pre-Clinical | |
| 6 | Inactivated | Inactivated + alum | Sinovac | Pre-Clinical | SARS |
| 7 | Inactivated | Inactivated | Beijing Institute of Biological Products/Wuhan Institute of Biological Products | Pre-Clinical | - |
| 8 | Live Attenuated Virus | Deoptimized live attenuated vaccines | Codagenix/Serum Institute of India | Pre-Clinical | HAV, InfA, ZIKV, FMD, SIV, RSV, DENV |
| 9 | Non- Replicating Viral Vector | MVA encoded VLP | GeoVax/BravoVax | Pre-Clinical | LASV, EBOV, MARV, HIV |
| 10 | Non- Replicating Viral Vector | Ad26 (alone or with MVA boost) | Janssen Pharmaceutical Companies | Pre-Clinical | Ebola, HIV, RSV |

*Note: *DRAFT landscape of COVID-19 candidate vaccines, URL: https://www.who.int/blueprint/priority-diseases/key-action/Novel_Coronavirus_Landscape_nCoV_Mar26.PDF; last accessed [26.03.2020]*

Fourthly, we see companies that possess a set of particular skills so they can be agile to speedily develop basically any kind of complex product, often possessing fairly flexible manufacturing facilities and equipment, but are often limited to low volume production. Examples include formula one teams, such as Williams Racing and McLaren that joined the UK ventilator challenge consortium [12].

All of these four types of companies typically own formal IP. These firms, being mostly large, commonly operate their own inhouse IP departments, thus possess awareness of IP relevance and understanding of how IP functions. However, their own IP which they developed before the start of the pandemic might not be exactly relevant in the CC-Sectors.

When entering CC-Sectors to support the scaling up of CC-Product manufacturing, these new entrants need to understand quickly how they can start manufacturing CC-Products in large volume. Essentially, to do this they have three options with regards to IP.

The first option is not to worry (and care) about incumbents CC-IP and go ahead, thus (wilfully) infringing existing CC-Products by reverse-engineering incumbent CC-Product. Under these circumstances, the owners of CC-IP may deny access to the proprietary information and know-how e.g., Italian 3D printing volunteers were denied access to proprietary information about ventilator valves forcing them to reverse-engineer the design [29] or in some cases enforce their IP and pursue the new entrants for infringement. So far this has been rare e.g. the case of Labrador Diagnostics LLC as previously mentioned, which is understandable during a pandemic due to potential reputational damages. However, new entrants choosing this option might become future targets for infringement claims from incumbents once the pandemic has ended. While new entrants are often large firms that can repurpose large manufacturing units, also well-established SMEs enter the production of CC-Products, e.g. local distilleries starting to produce sanitisers [24], [67]; 3D printing companies producing PPE [95]. Given their small scale, they might not be particularly prone to be attacked by incumbent IP owners but they would be in a vulnerable position if that situation did arise.

The second option for new entrants is to start designing CC-Products from scratch, possibly using their own engineering design competence and ability to procure rapid expert advice. For example, Dyson's ventilator design Covent, which was developed in partnership with MedTech consultancy. The Technology Partnership (TTP) [78]. Other efforts are being led by coalitions of large technology consultancies such as Cambridge Consultants, who typically work on medical device projects for clients but without holding their own IP [96]. However, starting from the ground up might not be the most efficient way to achieve impact during a pandemic as developing and obtaining medical approval for new CC-Product designs is likely to cause further delays. Choosing this option, new entrants would essentially develop novel CC-IP. However, if not carrying out careful freedom to operate analysis, they may infringe on existing background CC-IP owned by incumbents. Carrying out freedom to operate analysis may further delay the quick manufacturing of CC-Products. So new entrants are left balancing these risks under uncertainty and rapid changes to the technological, regulatory, economic and legal landscape.

The third option is to access CC-IP through teaming up with incumbents to produce some of the existing CC-Products manufactured prior to the pandemic by the incumbents only. For instance, the Smiths group and Penlon are incumbents with own ventilator designs who joined the UK ventilator challenge consortium [97]. These companies can then grant licenses and share CC-IP with consortium partners, including new entrants. Other incumbents, e.g. Medtronic, selectively pledged CC-IP for their Puritan BennettTM 560 - a basic ventilator model for which they have made available all designs and manufacturing details under a permissive license for a limited term [50]. For new entrants to access CC-IP this way is a way to avoid infringing existing IP owned by incumbents. For incumbents to pledge CC-IP is a way to facilitate the adoption of their technology during the pandemic, potentially with some lasting benefits beyond the pandemic. For instance, they can share CC-IP during a pandemic freely without charging any royalties using licenses that are time limited. If companies want to continue using that IP beyond the pandemic, the licensing terms would either prevent that or these companies would have to pay royalties.

Whether new entrants start developing designs for CC-Products from scratch, license existing CC-IP or infringe upon existing CC-IP, when scaling up production using their own resources new entrants are likely to develop novel (foreground) CC-IP. Given that they are faced with resource constraints having to manufacture CC-Products with equipment at their disposal and with materials they can access quickly through existing supplier relationships, they very well may end up adapting existing designs. This may lead them to find inventive designs or ways to e.g. manufacture CC-Products in a cheaper way. New entrants could possibly consider formally registering this new CC-IP, e.g. by filing patents.

Suddenly, incumbents may find themselves confronted with new entrants in their sector that infringe their CC-IP to some extent, which they find difficult to enforce during a pandemic, with new entrants developing subsequently their own CC-IP. Incumbents may fear that some of the new entrants will continue to stay in 'their' sectors i.e., CC-Sectors even after the pandemic eventually ends. Having established capabilities to manufacture CC-Products in innovative ways or using innovative designs e.g. low cost version for less developed countries, these new entrants may have few incentives to stop producing CC-Products and exit CC-Sectors. Incumbents may find themselves having helped to establish new competitors by not enforcing CC-IP during a pandemic. The extent to which this is a concern will depend on the incumbent's existing business model and IP strategy and their interaction strategy with the new entrants, therefore it is difficult to predict at present the severity of concern or whether it is justified.

## B. Scenario 2: Type 2 New Entrants - Voluntary Grassroot Initiatives, not-For Profit Organisations, Start-ups

During the past weeks we observed a large number of newly launched voluntary grassroots initiatives, not-for-profit organisations and start-ups joining the development and production of CC-Products. These initiatives often adopt explicitly or implicitly open source approaches widely sharing their designs. It seems we can distinguish two categories of such initiatives.

First, we have seen that a number of highly innovative voluntary initiatives got active to help with developing and manufacturing CC-Products. Those are typically founded by highly skilled people, such as engineers and scientists, and often develop fairly complex (high-tech) CC-Products, including hardware for ventilators, but also data platforms to collate pandemic data or tracking applications. For instance, a large number of institutions have developed complex epidemiological models, and geospatial models to understand the spread and development of the virus together with behaviour science, as well discuss the different approaches that have been used by governments (non-medical interventions) to slow the pandemic [32]. An important development is the utilisation of Artificial Intelligence (AI) methodologies, and specifically Deep Convolutional Neural Networks to detect COVID-19 from X-Ray images, as well as to identify the development and stage of the disease.

Quickly, several initiatives have come up with stunning new and affordable, easy to produce CC-Product designs which have often undergone highly sophisticated testing with state of the art equipment to which scientists have access through their labs. The initiatives then typically share their design drawings, CAD files as well as testing data through open source approaches, either through formal adoption of open source hardware and software licences or informally through statements of intent.

While these initiatives develop novel foreground CC-IP, they typically build on existing designs and by then adopting outbound licensing / open IP approaches [98], there is a potential (residual) risk that their designs may actually infringe upon existing CC-IP. As those initiatives usually have limited IP expertise and lack resources to access external legal support (e.g. patent attorneys), it is unlikely that most initiatives conduct freedom to operate analysis before launching their novel CC-Products. This lack of due diligence and IP clearance, with the open source initiatives probably excluding any liabilities and warranties for their designs, could lead any industrial adopter starting to produce an open source design in volume into trouble, suddenly unwillingly infringing incumbent CC-IP.

The second category includes the various initiatives that focus on the redesign or new manufacturing approaches for 'low-tech' CC-Products, i.e. with low technical and manufacturing complexity. Those CC-Products include face masks where we have seen numerous initiatives releasing patterns online calling for home production, i.e. crowd-manufacturing. Most of these initiatives might not be seen as highly innovative from an industrial standpoint as they produce fairly mature and 'dated' CC-Products, which might not be protected anymore by any alive IP. For instance, patents that once protected face mask designs may have long expired.

However, various initiatives focusing on low-tech CC-Products clearly innovate developing novel solutions that go beyond existing CC-Products. For instance, different initiatives have started to develop new face shield designs that are optimised for 3D printing. Those initiatives develop potentially patentable novel CC-IP, while not necessarily infringing incumbent owned CC-IP as this has probably already expired. Overall, those initiatives may not run into particular infringement risks, however, are likely to create novel foreground CC-IP. Given that most initiatives adopt open source licensing approaches they may not formally seek to protect their novel CC-IP through filing patents. However, we can possibly expect to see some trademark registrations appearing from some of the initiatives that become commercially viable, e.g. those also starting mass 3D printing production of face shields in Lithuania [99].

## C. Scenario 3: Incumbents

Certain expertise like vaccine and drug development are so unique to the incumbents in CC-Sectors that new entrants are less or almost unlikely to contribute significantly within the short time frames available during a pandemic. In these cases, incumbents in CC-Sectors developing new solutions or contributing to upscaling may end up infringing CC-IP of other incumbents. Our analysis shows that incumbents may take one or more of the following three approaches, each coming with different IP considerations.

One approach is the development of new technologies and solutions based on their pre-pandemic technologies and their expertise unique to CC-Sectors, thereby independently creating innovative novel CC-IP within their expertise. An example is the COVID testing developed by BioFire claimed to be based on its existing technologies (e.g. BioFire Filmarray). A second approach is that CC-Sector incumbents who were pre-pandemic component suppliers suddenly establish large scale manufacturing units to address supply-shortages. An example is INEOS, a chemical giant, supplier of one of the key ingredients used in sanitisers, established huge manufacturing capacities in France to produce about 1 million bottles per month. The third approach requires incumbents to collaborate by forming bilateral collaborations (e.g., agreement between Eli Lilly and AbCellera), establish new consortia (e.g., OPENCORONA consortium [84], COVID-19 Therapeutics Accelerator [82]) or joining existing networks (e.g., CEPI (Coalition for Epidemic Preparedness Innovations) and

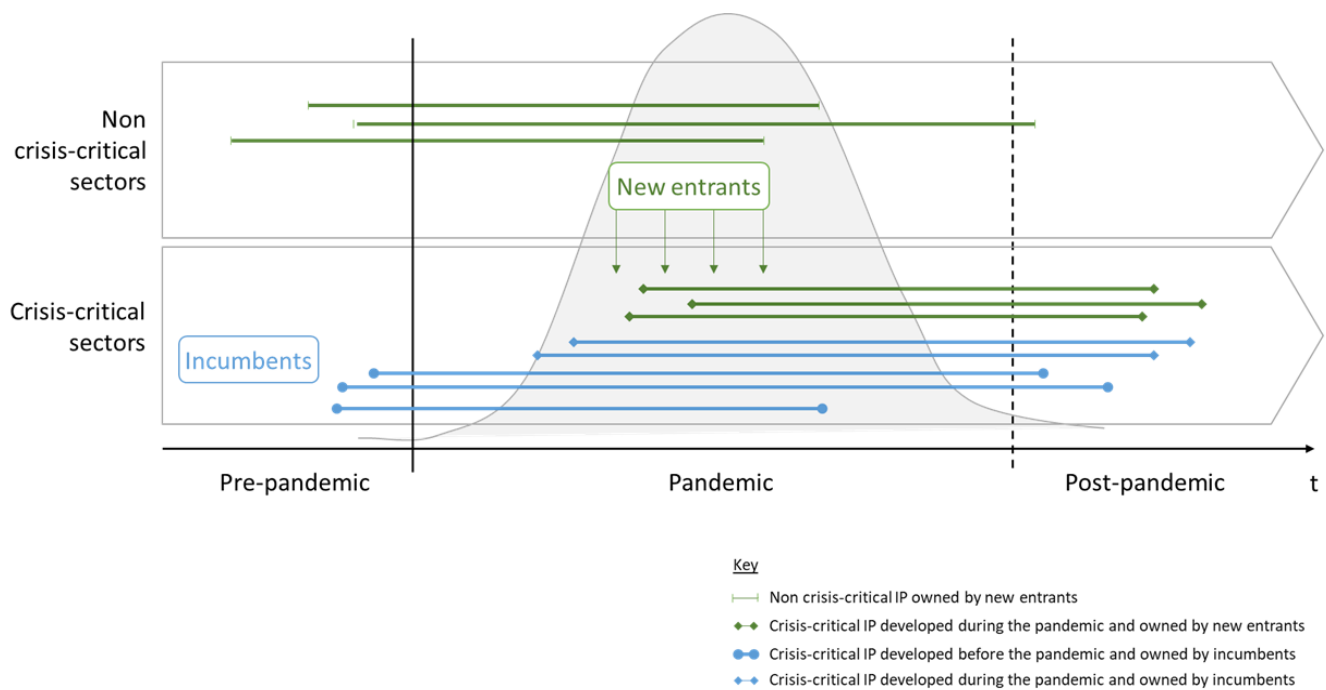

Fig. 4 Crisis-Critical IP (CC-IP) during a Pandemic

Europe's IMI (Innovative Medicines Initiative) [89]) and share IP among themselves to accelerate the efforts. The consortiums are likely to include non-commercial entities such as universities and research centres. Our analysis shows that consortiums at the prevention stage focus mainly on vaccine development. Examples include Horizon 2020's OPENCORONA consortium [84], consortium among Novartis, Bristol Myers Squibb and GSK [19], ChAdOx1 consortium involving universities and research centres. For treatment, consortiums are for CC-Products (e.g., UK Ventilator Challenge consortium [12]) and drugs (e.g., see [89] for several of R&D efforts by incumbents).

Two CC-IP considerations can be identified in the three approaches taken by incumbents. First, in the efforts by incumbents to accelerate CC-Product development, they may not conduct freedom to operate analysis as it is time taking. So, they may end up facing infringement lawsuits by other incumbents owning similar CC-IP (e.g., Labrador Diagnostics LLC's lawsuit against BioFire [100]). Second, when pre-pandemic suppliers establish mass production and manufacturing units, they might generate foreground IP during a pandemic, and would like to stay in the market post pandemic as well.

To summarise (Fig. 4), new entrants rushing to help the large scale production of high-tech CC-Products (e.g. ventilator production, diagnostic kits), particularly large ones entering CC-Sectors might be more at a risk to infringe active / alive CC-IP owned by incumbents than those entering low tech CC-Products, for which formally relevant IP might have already expired (e.g. face masks) or free alternatives exit. Overall, large firms might be more at risk to become targets for future (i.e.

after the pandemic) infringement claims than SMEs and voluntary initiatives. A particular consideration that should be mentioned is that while rushing into CC-Product manufacturing large firms may not perceive IP to be that urgent given a current crisis, which could turn out to be difficult in the long run as this could come with considerable IP risks. Incumbent owners of CC-IP might have a lower probability to sue voluntary initiatives, because they cannot claim damages, so could only ask for injunctions, which may cause themselves reputational damages. An exception might be the case of a CC-Product developed by a voluntary initiative, which is then produced in large volumes. However, a voluntary initiative giving away its IP open source would not make sufficient money to be a damage claim target. Rather those manufacturing an open source design without a clean freedom to operate situation could be targets for infringement claims. One also has to consider that new entrants, whether large manufacturing firms or voluntary initiatives, are likely to develop novel CC-IP during a pandemic, which they could use when continuing to stay in CC-Sectors.

## V. Approaches to Address Crisis Related IP Challenges

A pandemic is characterised as a crisis that calls for urgent, imminent large-scale action by governments, industrial players as well as a range of other societal actors. Governments take centre stage to orchestrate the rapid response to a pandemic. One of their many primary concerns is enabling the large scale production of Crisis-Critical Products (CC-Products), with the demand typically far exceeding the manufacturing capacities available from incumbent manufacturers in Crisis-Critical Sectors (CC-Sectors). This then calls for other firms to enter CC-Sectors and support the mass manufacturing of CC-Products by repurposing manufacturing lines. This leads to situations where manufacturing firms find themselves suddenly engaged in new relationships (e.g. UK ventilator challenge [22]), possibly even with companies that were competitors before the pandemic.

A pandemic, such as exemplified by the current COVID-19 crisis, also leads to wide ranging innovation activities, whether by incumbents or new entrants, being large firms themselves or various grassroot initiatives, not-for-profit organisations or start-ups etc. While governments' priority must be to enable the mass production of CC-Products, they should not forget to address potential IP concerns that incumbents or new entrants may have and think about ways to reduce IP related risks. At least three possible approaches are available to support the reduction of IP associated risks.

One legal approach that most governments have stems from the TRIPS agreement. Compulsory licensing is a tool that governments have in those countries that have adopted TRIPS [101]. It allows governments to use IP in crisis situations such as the COVID-19 pandemic. However, compulsory licenses are typically seen as a last resort measure. Compulsory licensing particularly helps governments to access and use CC-IP and thereby reduce IP associated risks mostly to new entrants, but are usually not favoured by incumbents, even though governments typically have to agree to pay a reasonable royalty for accessing their IP. In the current pandemic, countries that have already considered compulsory licensing with more to be expected include Chile, Canada and Israel [40], [41], [102].

Another, voluntary approach is to call owners of CC-IP (mostly incumbents) to pledge [48], [98], [103] their IP so that new entrants get non-exclusive licenses to use incumbents' CC-IP at least for the duration of the pandemic. Examples include firm specific pledges, e.g. by Medtronic [104], AbbVie [85] and SMITHS [105], but also the Open COVID Pledge (www.openCOVIDpledge.org) [52], [53] for industrially relevant IP as well as the Wellcome Trust pledge [36] for research publications and datasets. The pledge option, particularly with a time limited license, appears to be more friendly to incumbents while also de-risking IP challenges for new entrants. Pledges often also provide license templates that others can use and adjust to their needs.

Another approach to reduce IP associated risks and thus avoid any delays in fighting a pandemic are CC-IP pools, which can then be made available to a restricted group of companies (e.g., a consortium) only or to all interested firms that want to use that IP. A formal approach for governments would be to facilitate the development of patent pools [106], which have already been used in the pharmaceutical industry (e.g. Medicines patent pool).

## VI. CONCLUSIONS

Coronavirus is a virus family to which belongs the virus causing COVID-19. This paper shows that research and IP protections for coronavirus related inventions is not new. Patent protection for different forms of coronavirus already exist but not for the particular coronavirus type, SARS-CoV-2 that causes the COVID-19 disease. It is also evident that there is a time-lag between the outbreak and the materialisation of patents, and high number of citations to Non-Patent Literature, which shows the urgency of scientists for open data to put the information in the public domain.

What makes it difficult for IP to be given its required considerations during the early stage of a pandemic is the enormous sense of urgency which easily draws decision makers attention to huge and undoubtedly urgent operational challenges. With this paper we hopefully contribute to raising awareness that IP needs to be dealt with rather earlier than later during a pandemic in order to avoid that IP associated risks delay the mobilisation of resources so urgently needed for the research, development and mass manufacturing of Crisis-Critical Products.

This paper provides a terminology that helps to conceptualise IP considerations in times of a pandemic or global crisis that calls for urgent and large-scale actions from industrial stakeholders that suddenly find themselves engaged in new relationships that are associated with various uncertainties, not the least related to the use and sharing of IP with the particular problem that negotiating licensing agreements is typically time consuming.

From analysing the currently unfolding COVID-19 pandemic, we identify IP associated challenges relating to the prevention, diagnosis and treatment of a pandemic. We identify four types of stakeholders that are mostly concerned with IP considerations. These include governments (and inter-governmental organisations, such as the WHO, WIPO) who are called upon to orchestrate pandemic responses, incumbent manufacturing firms in CC-Sectors as well as new entrants that enter Crisis-Critical Sectors to assist incumbents. These new entrants can be manufacturing firms that have not produced Crisis-Critical Products prior to a pandemic, but also voluntary grassroot initiatives, start-ups, entrepreneurial scientists, etc.

We identify and analyse three scenarios in which different considerations around IP emerge. In the first scenario manufacturing firms enter Crisis-Critical Sectors to assist incumbents in mass manufacturing the volume required of Crisis-Critical Products. Those firms possess complementary capabilities and resources, so they can repurpose production lines. In the second scenario we discuss IP considerations when voluntary grassroot initiatives enter Crisis-Critical Sectors. In the third scenario we discuss R&D and manufacturing engagements and IP considerations for the incumbents in the CC-Sectors. We finally provide an initial discussion of three possible approaches to address IP concerns during a pandemic, namely compulsory licensing, IP pledges and IP pooling.

Crisis-Critical Intellectual Property: Considerations during the COVID-19 Pandemic © F. Tietze, P. Vimalnath, L. Aristodemou, J. Molloy 15

References# REFERENCES

AUTHORS

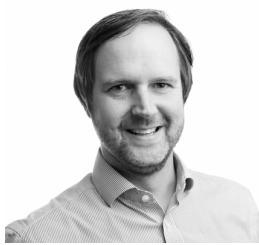

**Dr. Frank Tietze** is a Lecturer in Technology and Innovation Management at the University of Cambridge, Institute for Manufacturing. Within the Centre for Technology Management (CTM), he leads the Innovation and Intellectual Property Management (IIPM) Laboratory. He is a steering group member of Cambridge Global Challenges. His research has been published widely in leading international journals. Frank is departmental editor of IEEE Transactions on Engineering Management, editorial board member of World Patent Information, editorial review board member of LES Nouvelles, and the editor for the CTM working paper series. He is Affiliated to the Cambridge Centre for Intellectual Property and Information Law (CIPIL) and a member of the Innovation and IP research group at Chalmers University of Technology, Sweden.

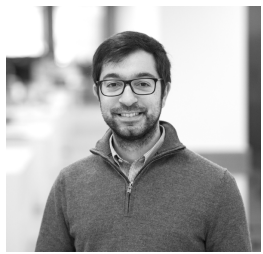

**Leonidas Aristodemou** is a Doctoral Researcher at the University of Cambridge.. He has been an enrichment scholar at The Alan Turing Institute, London, UK. He is supervised by Dr. Frank Tietze and advised by Prof. Tim Minshall. He obtained a Master's in Engineering (MEng) with Distinction, and a Bachelor of Arts (BA) in Engineering, from the University of Cambridge. He is an executive board member of the Cambridge University Engineers Association, a member of the EU AI Alliance, and a member of St. Edmund's College

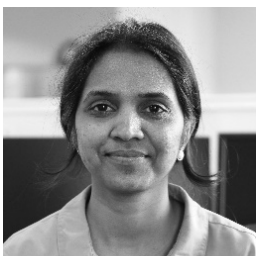

**Dr. Pratheeba Vimalnath** is a Post-Doctoral Research Associate at the University of Cambridge, Institute for Manufacturing. She is currently working for the project IPACST that aims to study the role of IP strategies in accelerating sustainability (social, environmental and economic) transitions. She is also a College Research Associate at the Wolfson College, Research Data Champion at the university's Data Champion programme, and an Interest Group Champion for the Cambridge Global Challenges (CGC). Prior to joining the IfM, she worked as an Oxford Martin Fellow at the University of Oxford to research on IP aspects of emerging open models for discovery and development of affordable medicines. She holds a doctorate degree in Intellectual Property Management, a Master's degree in Technology Management and a Bachelor's degree in Computer Science and Engineering, from India.

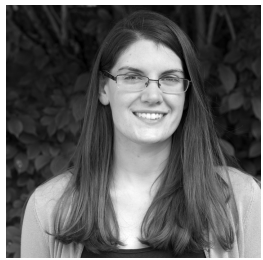

**Dr. Jenny Molloy** is a Shuttleworth Fellow at the University of Cambridge, studying the role and impact of open approaches to intellectual property for a sustainable and equitable bioeconomy. In particular she researches the potential for local, distributed manufacturing of enzymes to improve access and build capacity for biological research. This work combines technical development using synthetic biology-based platform technologies with qualitative research on challenges faces by molecular biologists globally through interviews and case studies.